\documentclass[reprint,onecolumn,PRF]{revtex4-2} 

\usepackage[utf8]{inputenc}
\usepackage{amsmath}

\usepackage{graphicx}
\usepackage{amsmath,amssymb,amstext} % Lots of math symbols and environments
\usepackage{enumerate}
\newcommand{\pderiv}[2]{\frac{\partial #2}{\partial #1} }
\newcommand{\deriv}[2]{\frac{\mathrm{d}#2}{\mathrm{d} #1} }

\newcommand{\Havg}[1]{\overline {#1}}

\newcommand{\Tzt}{\Theta_S}
\newcommand{\TSM}{\Theta_S}

\newcommand{\TB}{\Theta_B}

\newcommand{\TzH}{\tilde T_{\tilde z=\tilde H}}

\newcommand{\TO}{\tilde T_R}

\newcommand{\Qs}{\mathbf{q}_{Surf}}

\newcommand{\TMD}{\tilde T_{md} }

\newcommand{\delBL}{\delta_{BL}}

\newcommand{\Ra}{\hbox{Ra}_0}

\newcounter{magicrownumbers}

\begin{document}

\title{Atmospheric cooling of freshwater near the temperature of maximum density }
\author{Jason Olsthoorn}
\affiliation{Department of Civil Engineering, Queen's University, Kingston, Ontario, Canada, K7L 3N6 }
\email[]{Jason.Olsthoorn@queensu.ca}

\begin{abstract}
We perform three-dimensional direct numerical simulations of surface-driven convection near the temperature of maximum density $\tilde T_{md}$. A dynamic surface boundary condition couples heat flux through the surface to the induced convection, creating a dynamic equilibrium between the surface water temperature and the convection below. In this system, we identified three convective regimes: (1) free convection when the surface water temperature is above $\tilde T_{md}$, (2) penetrative convection when the surface water temperature is below $\tilde T_{md}$ and the convection is actively mixing the fluid layer, and (3) decaying convection when the convection weakens. We then predict the transitions between these regimes. Understanding these transitions is essential for the predicting timing of ice formation in natural systems.
\end{abstract}
\maketitle

% \begin{align}
%     \frac{1}{A} \deriv{t}{} \intV{\rho c_p T} & \\
%     &= -K_{Air} \left( \TzH - \TA \right)   + \sigma_{SB} \left( \epsilon \TA^4 - \epsilon_w \TzH^4 \right) ,\\
%     &= \left( -K_{Air}    + \frac{\sigma_{SB}\epsilon }{\left( \TzH - \TA \right)} \left( \TA^4 - \frac{\epsilon_w}{\epsilon} \TzH^4 \right)\right) \left( \TzH - \TA \right) ,\\
%     & \TzH^4 = \left(\TzH^\prime + \TA \right)^4 =  \TA^4 + 4 \TA^3 \TzH^\prime + \dots,\\
%     &= \left( -K_{Air}    + \frac{\sigma_{SB}\epsilon }{\left( \TzH - \TA \right)} \left( \TA^4 - \frac{\epsilon_w}{\epsilon} \left(\TA^4 + 4 \TA^3 \TzH^\prime\right) \right)\right) \left( \TzH - \TA \right) ,\\
%     &= - \left( K_{Air}    + 4 \epsilon_w \sigma_{SB}  \TA^3 \right) \left( \TzH - \TA \right)  + \sigma_{SB} \left(\epsilon - \epsilon_w \right) \TA^4,\\
%   Q = \rho_0 c_p \tilde F &=  \gamma \left( \TzH - \TO\right).
% \end{align}

\section{Introduction}

Recent work has highlighted that under-ice thermal stratification plays a key role for the biological dynamics in winter limnology \citep{yang_mixing_2020}. Moreover, this stratification, which persists through the whole ice-covered period, is largely determined by the thermal structure prior to the formation of lake ice \citep{yang_new_2021}. The two primary processes that control the pre-ice stratification are wind and surface-driven convection. This study is focused on the latter process to understand when convection, driven by atmospheric cooling, will shutdown, enabling the formation of ice.

To account for the heat loss through the air-water interface, 
\citet{hitchen_impact_2016} and others have demonstrated that the linearized surface boundary flux ($\Qs\cdot \mathbf{n}$) that accounts for longwave radiation, sensible heat flux, and evaporation depends upon the surface water temperature $\TzH$, and is written 
\begin{gather}
    \Qs\cdot \mathbf{n} =  \gamma \left( \TzH - \TO\right), 
    \label{eqn:Qs}
\end{gather}
with parameters $\gamma$ and $\TO$ that depend upon the physical parameters of the system, such as the air temperature, wind speed, and other meteorological variables above the water surface. \citet{olsthoorn_accounting_2023} then demonstrated that, provided that one takes into account the evolving surface boundary temperature, the magnitude of the heat flux through the water column is consistent with other convective systems.
% ~\citep[for example, see][]{clarte_effects_2021}
Further, there is an equilibrium surface temperature that results from a balance between the surface cooling that is driving the convection, and the convection-induced warming of the water surface. 

One limitation of the previous work of \citet{olsthoorn_accounting_2023} is that they assumed a linear equation of state. 
Many systems do not have a linear equation of state. In particular, the equation of state for freshwater is nearly quadratic about the temperature of maximum density ($\TMD \approx 4\ ^\circ \hbox{C}$). There has been significant recent research on convection with a nonlinear equation of state \citep[][and others]{wang_universal_2021,wang_how_2021,wang_penetrative_2019}. One such study by \citet{olsthoorn_cooling_2021} argued that convection with a quadratic equation of state has two distinct thermal regimes. In the first thermal regime, the bottom water temperature ($\TB$) decays exponentially towards $\TMD$ with a constant decay rate. Once the convection becomes weak enough, the second thermal regime evolves where the decay rate rapidly decreases. The transition between these two thermal regimes occurred when the ratio of nondimensionalized bottom to surface temperatures $\TB < -C_3\TSM$, for some constant $C_3$. However, \citet{olsthoorn_cooling_2021} used a fixed surface temperature, which does not account for the dynamic heat loss through the air-water interface.

This paper builds upon both \citet{olsthoorn_accounting_2023} and \citet{olsthoorn_cooling_2021} to study the induced convection as water approaches $\TMD$. To that end, we investigate convection with a quadratic equation of state using the linearized surface boundary condition \eqref{eqn:Qs}. We aim to determine: first, how rapidly does the water temperature decrease and second, when does convection shut down in this system. 

The remainder of this paper is organized as follows: we continue in \S\ref{sec:methods} with a discussion of the numerical methods and specific cases performed. We then discuss each of the objectives in turn in \S \ref{sec:decay}--\ref{sec:TS}, before finally concluding in \S \ref{sec:conclusion}

\section{Problem Setup} \label{sec:methods}

\begin{figure}
    \centering
    \includegraphics[width=\textwidth]{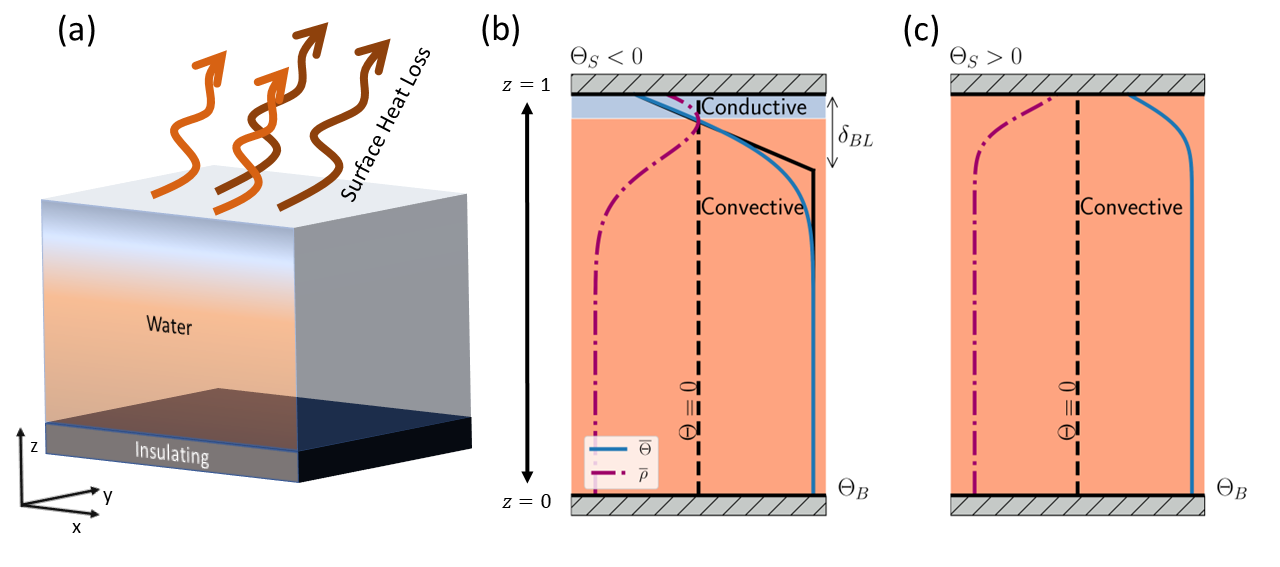}
    \caption{(a) A schematic of the numerical domain. Diagram of the thermal structure within the domain for (b) $T_S<0$ and (c) $T_S>0$. An approximate piecewise-linear temperature profile is illustrated as a black line in (b). This figure is adapted from \citet{olsthoorn_cooling_2021}. }
    \label{fig:Diagram}
\end{figure} 

We consider direct numerical simulations of a three-dimensional fluid domain with an initial uniform water temperature above $\TMD$. The surface temperature boundary condition is given in \eqref{eqn:Qs}, and the bottom boundary condition is insulating. Simulations were performed with the spectral solver SPINS \citep{subich_simulation_2013}, using pseudospectral spatial derivatives and a third-order time-stepping scheme. The horizontal domain is periodic and a Chebyshev grid is used in the vertical to resolve the top thermal boundary layer. Due to a limitation of the numerical code, both the top and bottom boundary conditions are no-slip. However, we do not believe that this will substantially affect our conclusions as our results from Regime I ( when the surface temperature is above $\TMD$) are similar to those found in \citet{olsthoorn_accounting_2023}, which use a free-slip surface velocity condition, and \citet{clarte_effects_2021}. Figure~\ref{fig:Diagram}a is a diagram of the problem setup.

Similar to the work of \citet{olsthoorn_cooling_2021}, we nondimensionalize the system by considering the domain depth $H$, the temperature difference $\Delta \tilde T = \TMD - \TO$, the timescale of thermal diffusion $\tau_\kappa=\frac{H^2}{\kappa}$ and the corresponding velocity scale $\frac{H}{\kappa}$, where $\kappa$ is thermal diffusivity. 
The equations of motion for nondimensional velocity $\mathbf{u}$ and temperature $\Theta$ are then written
\begin{gather}
	\left( \pderiv{t}{} + \mathbf{u} \cdot \nabla \right)\mathbf{u} = - \nabla P  +  \Ra \hbox{ Pr } \Theta^2 \mathbf{\hat k} + \hbox{Pr} \nabla^2 \mathbf{u},
	\label{eqn::momentum}\\
	\left( \pderiv{t}{} + \mathbf{u} \cdot \nabla \right) \Theta = \nabla^2 \Theta,  \qquad 
	\nabla \cdot \mathbf{u} = 0. \label{eqn::DivFree}
\end{gather}
We have defined the Rayleigh number ($\Ra$), Prandtl number (Pr), and nondimensional temperature ($\Theta$) as 
\begin{gather}
\Ra =  g C_T \Delta \tilde T^2\frac{H^3}{\kappa \nu}, \qquad \hbox{Pr} = \frac{\nu}{\kappa}, \qquad \Theta = \frac{\tilde T - \TMD}{\Delta \tilde T}, 
\end{gather}
where $\nu$ is the molecular viscosity of water, $C_T$ is the thermal expansion coefficient for a quadratic equation of state, and $g$~is the gravitational acceleration. We fixed Pr=9. In this scaling, $\TMD$ corresponds to $\Theta=0$. 
Further, we can then rewrite the surface boundary conditions as
\begin{gather}
    \pderiv{z}{\Theta} = -\beta \left( \Theta + 1\right),\qquad z = 1 \label{eqn::TopBC},
\end{gather}
with the Biot number $\beta$, 
\begin{gather}
    \beta = \frac{\gamma H}{c_p \rho_0 \kappa}, 
\end{gather}
where $c_p$ is the heat capacity of water and $\rho_0$ is a reference density. The Biot number $\beta$ controls the relationship between the surface temperature and the convective flux through the water. We can see this relationship by re-arranging equation \eqref{eqn::TopBC} as
\begin{gather}
    \Theta = -1 - \frac{1}{\beta} \pderiv{z}{\Theta}, \qquad z = 1.\label{eqn::TS}
\end{gather}
% where $\Tzt = \Theta(z=1)$ is the surface temperature. 
We find that the work of \citet{olsthoorn_cooling_2021}, which considered a fixed surface temperature condition, reduces to the limiting cases where $\beta \to \infty$. Conversely, for finite values of $\beta>0$, the surface temperature is modified by the convective heat flux to the water surface. As the strength of the convection is determined, in part, by the surface temperature, the system naturally tends towards a dynamic equilibrium between the surface and bottom water temperatures. 

% \subsection{Heat Fluxes}

In \citep{olsthoorn_cooling_2021}, the Authors showed that when the mean surface water temperature ($\TSM$) decreases below $\TMD$ ($\Tzt<0$), there exists a stable (conductive) layer at the upper boundary, above a convective layer below (see Figure~\ref{fig:Diagram}b). This is a result of the quadratic equation of state where the decreasing temperature with height results in a decreasing density profile near the surface. However, for finite $\beta$, the surface temperature depends upon the flux of heat to the surface and can result in a surface temperature above $\TMD$ ($\Tzt>0$). In this case, the entire domain is unstably stratified (see Figure~\ref{fig:Diagram}c). This permits a free-convection regime without any stable layer. 
% The Biot number $\beta$ determines whether the surface temperature $\TSM$ is above or below $\TMD$ ($\TSM =0$) for a given convective heat flux (see \eqref{eqn::TS}).  

Based upon the previous work of \citet{olsthoorn_accounting_2023} and \citet{olsthoorn_cooling_2021}, we expect there to exist three different convective regimes in this system:
\begin{enumerate}[(I)]
    \item $\TSM>0$ - \textbf{Free-convection} without an upper conductive layer
    \vspace{-6pt}
    \item $\TSM<0$ and $\TB>-C_3 \TSM$ - \textbf{Penetrative-convection} with an upper conductive layer and strong convection below,
    \vspace{-6pt}
    \item $\TSM<0$ and $\TB<-C_3 \TSM$ - \textbf{Decaying-convection} with an upper conductive layer and weak/decaying convection below.
\end{enumerate}

As we will see below, it is possible for the convective system to progress through each of these regimes as the mean water temperature decreases. 

\subsection{Simulated Cases}

We performed a series of numerical simulations with a quadratic equation of state at different Rayleigh numbers and $\beta$ values (see Table \ref{Table::NumParams}). We initially perturb the three velocity components with a random perturbation sampled from a Normal distribution scaled by $10^{-2}$. The numerical resolution (Nx $\times$ Ny $\times$ Nz) is similar to those of \citet{olsthoorn_cooling_2021} and \citet{olsthoorn_accounting_2023}, such that $\max  \Delta \mathbf{x}<3\eta_B$, where we compute the Batchelor scale 
$\eta_B = \left(\Havg{\varepsilon}\right)^{-\frac{1}4} \Pr^{-\frac{1}2}$, 
for horizontally averaged viscous dissipation rate $\Havg{\varepsilon}$ and grid spacing $\Delta \mathbf{x}$. Similarly, we ensure that $N_B \ge 10$ grid points over the thermal boundary layer, defined by $\delta_{BL} = \frac{\TSM - \TB}{\pderiv{z}{\Theta}(z=1)}$. Table values are reported after the initial instability and the system has reached quasi-steady state with a finite $\delta_{BL}$. 

The domain width (Lx), depth (Ly), and height (Lz) were selected to have a large aspect ratio of four. At higher $\Ra$, limited computational resources forced us to reduce the aspect ratio to two. The results are consistent between the different $\Ra$ cases.

\begin{table}
	\begin{center}
		\begin{tabular}{ccccccc}
			Case & Ra & $\beta$ & Resolution  & Domain Size  & max $\Delta \mathbf{x}/\eta_B$ & $N_B$ \\
			 &  &  & (Nx $\times$ Ny $\times$ Nz)   &   (Lx $\times$ Ly $\times$ Lz) &  &  \\
    			\hline
			1 & $9 \times 10^5$ & 1 & 128$\times$128$\times$128 & 4$\times$4$\times$1 & 1.472 & 20 \\
			2 & $9 \times 10^5$ & 10 & 256$\times$256$\times$256 & 4$\times$4$\times$1 & 1.098 & 37 \\
			3 & $9 \times 10^5$ & 25 & 256$\times$256$\times$256 & 4$\times$4$\times$1 & 1.165 & 36 \\
			4 & $9 \times 10^5$ & 100 & 256$\times$256$\times$256 & 4$\times$4$\times$1 & 1.190 & 37 \\
			5 & $9 \times 10^5$ & 1000 & 256$\times$256$\times$256 & 4$\times$4$\times$1 & 1.163 & 38 \\
			\hline 
			6 & $9 \times 10^6$ & 1 & 128$\times$128$\times$128 & 4$\times$4$\times$1 & 2.800 & 17 \\
			7 & $9 \times 10^6$ & 10 & 512$\times$512$\times$256 & 4$\times$4$\times$1 & 1.206 & 29 \\
			8 & $9 \times 10^6$ & 25 & 512$\times$512$\times$256 & 4$\times$4$\times$1 & 1.406 & 28 \\
			9 & $9 \times 10^6$ & 100 & 512$\times$512$\times$256 & 4$\times$4$\times$1 & 1.565 & 28 \\
			10 & $9 \times 10^6$ & 1000 & 512$\times$512$\times$256 & 4$\times$4$\times$1 & 1.662 & 27 \\
			\hline 
			11 & $9 \times 10^7$ & 25 & 512$\times$512$\times$256 & 2$\times$2$\times$1 & 1.592 & 21 \\
			12 & $9 \times 10^7$ & 100 & 512$\times$512$\times$256 & 2$\times$2$\times$1 & 1.918 & 20 \\
			\hline &  &  &  &  &  &  \\
		\end{tabular}
	\end{center}
	\caption{Table of the parameters for each numerical simulation. Note that the domain width for $\Ra=9\times10^7$ is half of the other cases. We report the maximum grid spacing relative to the Batchelor scale ($\eta_B$) defined based upon horizontally averaged dissipation rates. We further report the number of grid points within the surface boundary layer ($N_B$) defined based upon the surface temperature gradient. Both $\eta_b$ and $N_B$ are calculated after the initial onset of convection.}
	\label{Table::NumParams}
\end{table}

\section{Results} 

\begin{figure}
    \centering
    \includegraphics[width=1.0\textwidth]{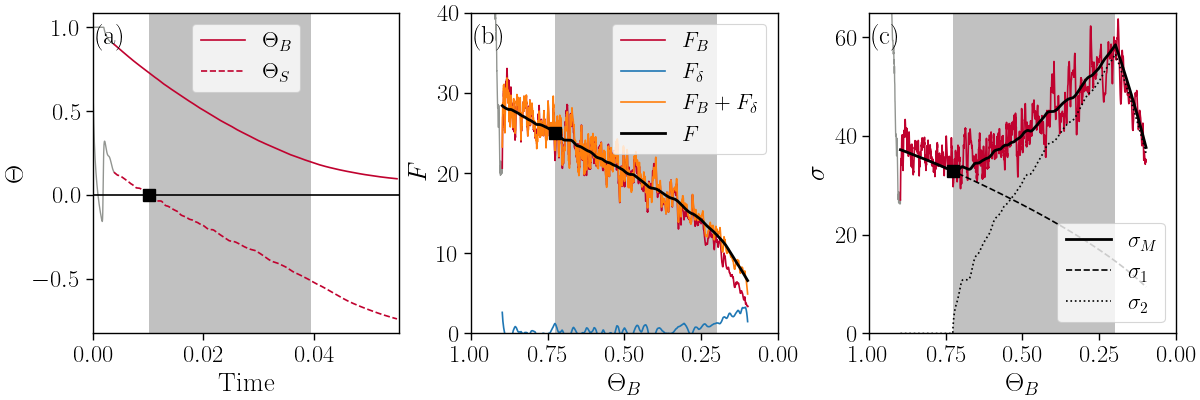}
    \caption{(a) A plot of the time evolution of mean bottom water temperature $\TB$ and surface water $\TSM$. Here and elsewhere, the initial transient is included as light-grey line. (b) The simulated surface heat flux ($F$) as a function of bottom water temperature $\TB$. (c) The temperature decay rate $\sigma$ as a function of $\TB$. A black dot denotes where $\TSM=0$, and the penetrative-regime ( Regime II) is denoted by the grey banner.    }
    \label{fig:FluxDecomposition}
\end{figure}

After an initial transient, the surface-driven convection mixes the interior water to depth (see Figure~\ref{fig:Diagram}b,c). We denote this near-uniform bottom water temperature $\TB$, which we calculate as a mean temperature below the mid-depth. Figure~\ref{fig:FluxDecomposition}a is a plot of the mean surface temperature $\TSM$ and bottom temperature $\TB$ for Case~8 ($\Ra=9.0\times 10^6, \beta=25$). This case is presented as it demonstrates the three thermal regimes listed above. The convection initially drives the surface water above $\TMD$ ($\TSM>0$). However, as $\TB$ decreases, the convection weakens and the surface temperature drops below zero near $t\approx 0.01$. The penetrative-convection continues until $\TB$ approaches zero and the decaying-convection regime starts around $t\approx 0.04$. 

The rate at which the water temperature ($\TB$) decreases is determined by the flux of heat through the water surface ($F$). Similar to \citet{olsthoorn_cooling_2021}, we model the mean temperature profile as piecewise linear (see Figure~\ref{fig:Diagram}b). Computing the time-derivative of this profile, we can show that
\begin{gather}
F \approx -\underbrace{\deriv{t}{\TB}}_{F_B}  - \underbrace{\deriv{t}{}  \left(\left(\frac{\TSM - \TB}{2}\right)\delBL\right)}_{F_\delta} .\label{eqn::F}
\end{gather}
The surface cooling is the sum of two components: the cooling of the bottom water ($F_B$) and the growth of the thermal boundary layer ($F_\delta$). 
At high $\Ra$, when the convection is vigorous, the boundary layer thickness is small and the surface flux is dominated by decreasing $\TB$ ($F\sim F_B, F_B\gg F_\delta$). When convection weakens ($\TB\to0$), as in the decaying-convection regime, a significant portion of the surface heat loss is used to grow the top boundary layer ($F_\delta>0, F_\delta \sim F_B$). Looking at Figure~\ref{fig:FluxDecomposition}b, we observe a growth in $F_\delta$ when $\TB \approx 0.2$, which, as we discuss below, occurs at the transition to the decaying-convection regime.

\subsection{How rapidly does the water temperature decrease?} \label{sec:decay}

We can quantify how quickly the mean water temperature decreases through the decay rate $\sigma$, which quantifies the rate at which the temperature approaches an equilibrium value. 
When $\TSM>0$, convection will drive the mean water temperature to the mean surface temperature ($\TB\to\TSM$). However, when $\TSM<0$, convection will only decrease the mean water temperature to the temperature of maximum density ($\Theta\to0$). We quantify the rate at which $\TB$ approaches equilibrium as the decay rate 
\begin{gather}
    \sigma = \frac{-1}{\TB - \Theta_0} \deriv{t}{\TB}, 
\end{gather}
where $\Theta_0 = \max\{\TSM,0\}$. As can be shown from \eqref{eqn::F}, where $F_\delta\ll1$, $\sigma$ is equivalent to the Nusselt number. Thus, as suggested by \citet{olsthoorn_accounting_2023} and others, we would expect that for free convection, $\sigma$ should scale with the surface-temperature-dependant Rayleigh number Ra$_e = \Ra \left( \TB - \Theta_0\right)^2$,
\begin{gather}
    \sigma_1 \sim C_1 \hbox{Ra}_e^p  , \qquad \TSM>0. 
\end{gather}
However, the work of \citet{olsthoorn_cooling_2021} suggests that for $\TSM<0$, two different thermal regimes are possible with different decay rates:
\begin{gather}
\sigma_2 \sim  C_2 \left( \Ra \TSM^2  \right)^p \begin{cases}
1, &  \TB \ge -C_3 \TSM \\
 \left(\frac{\TB}{C_3 \TSM}\right)^{2p}, &  \TB < -C_3 \TSM
\end{cases}, \qquad \TSM<0
, \label{eqn::SigmaScl}
\end{gather}
where we have modified the model of \citet{olsthoorn_cooling_2021} to account for the changing surface temperature.

Figure~\ref{fig:FluxDecomposition}c is a plot of the numerically computed $\sigma$ compared with $\sigma_1$ and $\sigma_2$. Similar to the values of \citep{olsthoorn_cooling_2021,olsthoorn_accounting_2023}, we fit $C_1 = 0.28, p=0.31, C_2 =0.6, C_3 = 0.385$. We find that $\sigma_1$ agrees well with the simulated $\sigma$ during the free-convection regime $\TSM>0$ and that $\sigma_2$ agrees well during the decaying-convection regime, but neither agree well during the penetrative-convection regime (see the grey shaded region in Figure~\ref{fig:FluxDecomposition}). 

We can approximate the decay rate over the entire simulation by considering both contributions to the temperature decrease as 
\begin{gather}
    \sigma_M = \sqrt{\sigma_1^2 + \sigma_2^2}, \label{eqn:sigM}
\end{gather}
which is essentially the norm of two orthogonal contributions to the temperature decay rate. We find good agreement between this combined model ($\sigma_M$) and the simulated temperature decay ($\sigma$) across all three temperature regimes (Figure~\ref{fig:FluxDecomposition}c).

\begin{figure}
    \centering
    \includegraphics[width=0.9\textwidth]{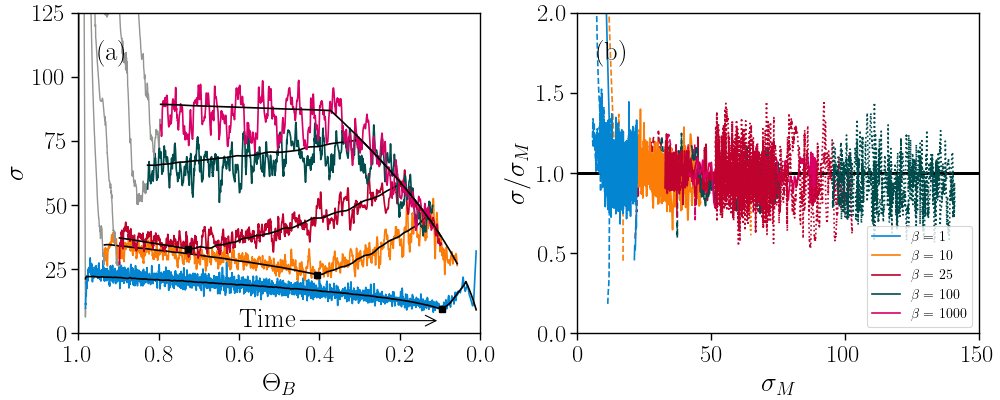}
    \caption{(a) Plot of the compute decay rate $\sigma$ vs. $\TB$ for the cases $\Ra=9\times10^{6}$. The combined model $\sigma_M$ is included as a black line. (b) Comparison of the computed $\sigma$ vs. $\sigma_M$ for all simulated cases. The model $\sigma_M$ is given in \eqref{eqn:sigM} with fit parameters $C_1 = 0.28, p=0.31, C_2 =0.6, C_3 = 0.385$. Note that data from $\TB<0.01$ was removed.   }
    \label{fig:enter-label}
\end{figure}

Further this combined model agrees with all of the simulated cases (See figure~\ref{fig:enter-label}). Despite the complexity of this model, and the significant oscillations, this model captures the temperature decay rate through all three identified thermal regimes.

% \subsection{When does the surface water temperature reach $\TMD$ ($\TSM=0$)} \label{sec:TS}
\subsection{When do the regime transitions occur? } \label{sec:TS}

It is import to note that the decay rate $\sigma$ is itself a function of both $\TB$ and $\TSM$. Thus, we need additional information to predict $\TSM$ given measurements of $\TB$ (or vice versa). We have shown (Figure~\ref{fig:FluxDecomposition}b) that except during the decaying-convection regime (where the boundary layer grows relatively rapidly), the surface flux is well approximated by $F \approx -\sigma_M\left(\TB - \Theta_0\right)$. Further, following the approach of \citet{olsthoorn_accounting_2023}, we can use the surface boundary condition~\eqref{eqn::TopBC} to determine 
\begin{gather}
    \sigma_M(\TB,\TSM; \Ra) \left(\TB - \Theta_0\right) = \beta \left( \TSM + 1\right). 
\end{gather}
We further isolate a factor of $\Ra^p$ from $\sigma_M$ to rewrite 
\begin{gather}
   \frac{\Ra^p}{\beta}  = \frac{1}{ \Gamma_M(\TB,\TSM)}  \left( \frac{\left( \TSM + 1\right)}{\left(\TB - \Theta_0\right)}  \right), \qquad  \Gamma (\TB,\TSM) = \frac{\sigma_M(\TB,\TSM; \Ra)}{\Ra^p}. \label{eqn:ParameterEstimate}
\end{gather}
The scaled decay rate $\Gamma_M$ is independent of the Rayleigh number. That is, for any value of $\TSM$, we have identified an implicit equation to determine $\TB$ (or vice versa), with the key parameter $\Ra^p/\beta$. 

\begin{figure}
    \centering
    \includegraphics[width=0.9\textwidth]{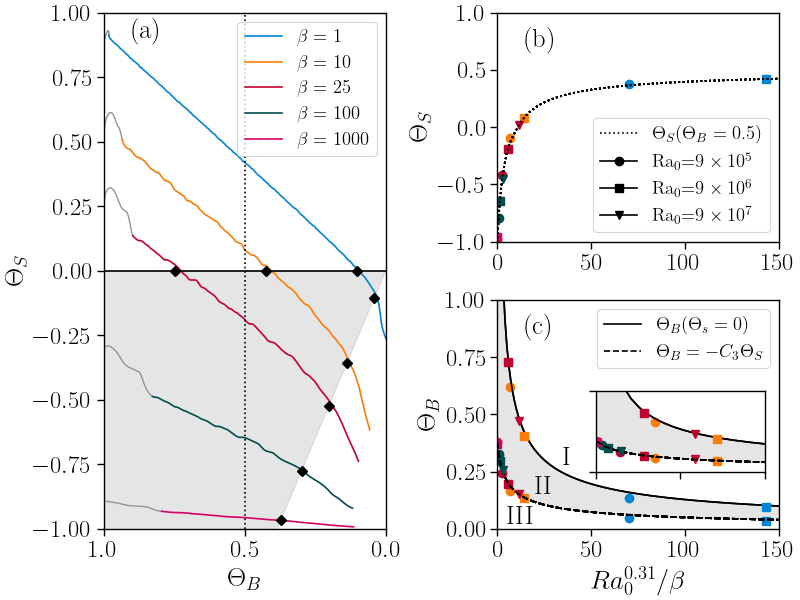}
    \caption{(a) Surface temperature ($\TSM$) vs. bottom temperature ($\TB$) for $\Ra=9\times10^6$. Black diamonds denote the predicted regime transitions for each case when $\TSM=0$ and when $\TB = -C_3 \TSM$. (b) The surface temperature $\TSM$ for a picked value of $\TB=0.5$ (corresponding to the vertical dotten line in (a)). (c) The bottom temperature $\TB$ at the regime boundaries ($\TSM=0$ and $\TB = -C_3 \TSM$) as a function of $\Ra^p/\beta$. 
    % The horizontal (vertical) lines correspond to a fixed value of $\TSM$ ($\TB$). 
    The regimes are denoted (I) free-convection regime, (II) penetrative-convection regime [shaded grey], and (III) decaying-convection regime. The inset presents the same data on a reduced x-axis of $\Ra^{0.31}/\beta$=[0,20].  }
    \label{fig:SurfaceTemp}
\end{figure}

We demonstrate the validity of this model by investigating the evolution of $\TSM$ as a function of $\TB$. Figure~\ref{fig:SurfaceTemp}a is a plot of $\TSM$ as a function of $\TB$ for different values of $\beta$ with $\Ra=9\times10^6$. For a given value of $\TB = 0.5$, we find good agreement between the predicted values of $\TSM$ from \eqref{eqn:ParameterEstimate} and those found in the simulations (Figure~\ref{fig:SurfaceTemp}b). 
% We note that we do not expect such good agreement for during the decaying-convection regime, where $\TB < -C_3 \TSM$, as $F_\delta$ is not negligible during that regime.   

We have identified three key convective regimes in this system. The regime transition between Regime I (free-convection) and Regime II (penetrative-convection) occurs when $\TSM=0$; when the surface temperature equals the temperature of maximum density. The second regime transition between Regime II (penetrative-convection) and Regime III (decaying-convection) occurs when $\TB=-C_3 \TSM$; when the convection significantly weakens and the thermal boundary layer grows rapidly. Given the physical parameters of this system, we can use \eqref{eqn:ParameterEstimate} to determine the critical values of $\TB$ when these transitions are expected to occur. Figure~\ref{fig:SurfaceTemp}c includes a comparison of the predicted and simulated points of transition. We find that the model is reasonable at predicting these transitions, though we note that the value of $\TB$ for $\TSM=0$ is slightly under predicted.

\section{Discussion and Conclusion} \label{sec:conclusion}

The dynamic surface boundary condition \eqref{eqn::TopBC} prescribes how the heat flux through the surface is coupled to the surface water temperature. In a cooling system, like the one described here, the atmospheric forcing cools the surface, which drives convection. However, the induced convection in the water results in a net-positive vertical heat flux, which warms the surface from below, increasing the surface temperature. That is, there is a dynamic equilibrium for the surface temperature as it both generates, and is modulated by, the convection.

The boundary condition \eqref{eqn::TopBC} can be viewed as a parameterization for the finite conductivity of the atmospheric boundary layer above the water. This is similar to the significant body of literature discussing the role of the finite conductivity of the bounding plates used in experimental studies of Rayleigh-Bénard convection. For example, \citet{chilla_ultimate_2004} argue it is the finite conductivity of the bounding plates that explain why the `ultimate'-convective regime is not found in certain experiments. \citet{verzicco_effects_2004} and \citet{brown_heat_2005} determine correction factors to account for this finite conductivity, assuming a linear equation of state.  \citet{wittenberg_bounds_2010} further discusses the role of $\beta$ on heat flux bounds across the whole liquid/solid-boundary system. This present paper focuses on the role of the nonlinear equation of state in modifying the temperature evolution of the system.  

The present manuscript builds upon these previous studies to investigate convection with a quadratic equation of state and has revealed three convective regimes: Regime I free-convection, Regime II penetrative-convection, and Regime III decaying-convection (See figure~\ref{fig:FluxDecomposition}). The boundaries between these regimes are then determined where $\TSM=0$ (Regime I to Regime II) and where $\TB=-C_3 \TSM$ (Regime II to Regime III). The critical values of $\TB$ can be estimated from \eqref{eqn:ParameterEstimate}. For a known $\TB$, such as from field measurements, \eqref{eqn:ParameterEstimate} can be further used to estimate the corresponding surface temperature.

One of the aims of this work is to determine when we expect to find ice-formation as a function of $\TB$. These direct numerical simulations are not capable of including a freezing surface flux, but, as shown by \citep{wang_how_2021}, the shape of the ice is determined by the water convection. Future work would include an ice-growth model to predict the timing and growth of ice formation. That said, in the absence of external forcing, we would expect that the water surface will start to freeze once its temperature is cooled below the freezing point. Due to the rapidly decreasing surface heat flux, we hypothesize that ice formation will occur shortly after the transition from regime (II) to regime (III). Future work will investigate how these regimes would be updated in the presence of wind-driven mixing, and if the regime transitions can be identified in field measurements. 

\section*{Acknowledgements}
We want to thank Dan Robb and Ted Tedford for their feedback on this work. This work was funded in part by the Natural Sciences and Engineering Research Council of Canada. Computational support was provided by the Queen's University Centre for Advanced Computing.

\bibliographystyle{abbrvnat}
\bibliography{references}

\end{document}